\documentclass[a4paper,12pt]{revtex4}
\usepackage{amsmath}  
\usepackage{amssymb}  
\usepackage{dsfont}   
\usepackage{fixmath} 

\begin{document}
\title{Quantum stochastic processes in two dimensional space-time}
\author{Alberto C. de la Torre}\email{delatorre@mdp.edu.ar}

\affiliation{Universidad Nacional de Mar del Plata}

\begin{abstract}
Several stochastic processes with virtual particles in two
dimensional space-time are presented whose mean field equations
coincide with Schr\"{o}dinger, Dirac, Klein-Gordon and the quantum
mechanic equation for a photon. These processes could be used to
detect discrete space-time features at the Planck scale.
\end{abstract}
\maketitle

\section{Introduction}
With the appearance of interference effects in the early days of
quantum mechanics it became clear that the dynamic of quantum
systems could not be simulated by particles with \emph{classical}
properties such as definite position and momentum, point
interaction, sharp energy, etc. and the \emph{wave} representation
was preferred. However with the advent of quantum field theory the
particle picture reappeared, not as classical entities, but as
virtual particles with ephemeral existence, created and annihilated
at all space-time points and with all energy and momentum values not
satisfying the usual energy-momentum relations, that is, \emph{``off
the mass shell''}. This theory allows for an interpretation of
quantum mechanics where the virtual particles are assigned some
ontological reality as components building the quantum
field\cite{dlt10}. In this interpretation, Feynman diagrams are not
only a mathematical term of a perturbation expansion but represent
really existent processes.

Historically, virtual particles appeared when the solutions of the
dynamic equations of quantum mechanics were expanded in ``normal
modes'' that were interpreted as quantum excitations. In this work
we take the opposite way, that is, we assume some types of virtual
particles with some type of interaction an we show that their
collective behaviour is described by the well known dynamic
equations of quantum mechanics. For simplicity, and considering
possible computer simulations, these  ``quantum stochastic
processes'' are studied in two dimensional space-time that include
the essential physical features. Furthermore we concentrate on free
systems, that is, without external sources.

The usual quantum mechanic equations are presented as arising from
different energy-momentum relations, including the less well known
quantum mechanic equation for the photon. These are specialized for
two dimensional space-time and several stochastic processes are
presented whose mean field equations coincide with quantum mechanic
ones. Finally some speculations concerning possible interesting non
equilibrium effects of the processes are suggested.
\section{Quantum dynamics for free systems}
The energy-momentum relation of a free physical system determines
its dynamics because energy and momentum are the generators of space
and time translations. For nonrelativistic systems the
energy-momentum relation is
\begin{equation}\label{epSchro}
    E=\frac{P^{2}}{2m}
\end{equation}
where $E$ is the energy and $P^{2}= \mathbf{P\cdot P} $ is the
momentum amplitude squared. In the relativistic case the relation is
(in units such that $c=1$ and with signature $+---$)
\begin{equation}\label{epKG}
    p^{\mu} p_{\mu} =E^{2}- P^{2}= m^{2}\ .
\end{equation}
Some inconveniences due to the \emph{quadratic} nature of the
expression above are avoided by postulating a \emph{linear}
energy-momentum relation, that must \emph{also} satisfy the
relativistic quadratic relation. This can only be achieved with
coefficients $\gamma^\mu$ satisfying a Clifford algebra $
\{\gamma^\mu,\gamma^\nu\} = 2 \eta^{\mu\nu} $. Then we have
\begin{equation}\label{epDirac}
 \gamma^\mu p_{\mu} = \gamma^{0}E- \mathbold{\gamma}\cdot\mathbf{P}= m\ .
\end{equation}

The relativistic relation (\ref{epKG}) is also valid for massless
particles, $m=0$, and in this case the energy becomes a linear
dependence with the momentum amplitude
\begin{equation}\label{epMw0}
    E= |\mathbf{P}|\ .
\end{equation}
For massive particles, spin $\mathbf{S}$ and momentum $\mathbf{P}$
are decoupled and can have arbitrary directions; however, for
particles with zero mass, like the photon, special relativity
requires that spin and momentum must be collinear (transversality
constraint), that is,
\begin{equation}\label{transv}
    \mathbf{S\times P}=0\ ,
\end{equation}
and the sign of $\mathbf{S\cdot P}$ corresponds to the positive or
negative helicity of the photon. The absolute value in
Eq.(\ref{epMw0}) is problematic when we transform the observables
into operators. In order to avoid the absolute value, it is
convenient to combine the linear dependence of energy and momentum
with the two possible helicity states of the photon by choosing the
energy-momentum relation for the photon as\cite{dltPhoton}
\begin{equation}\label{epMw}
    E= \mathbf{S\cdot P}\ .
\end{equation}
When this quantity is negative, it must be interpreted as a photon
with \emph{positive} energy but with negative helicity.

In quantum physics, the energy and momentum observables are
represented by operators in a Hilbert space and when we choose for
it the space of square integrable functions, where $E\rightarrow
i\partial_{t}$ and $\mathbf{P}\rightarrow -i\nabla$ (in units such
that $\hbar=1$), we obtain the well known Schr\"{o}dinger, Klein-Gordon
and Dirac equations. In the representation for spin 1 where the spin
operators $S_{l}$ have matrix elements
$(S_{l})_{jk}=-i\varepsilon_{ljk}$ the quantum mechanics equation
for the photon, obtained from Eq.(\ref{epMw}), is\cite{dltPhoton}
\begin{equation}\label{QMequPhoton}
 i\partial_{t}\Psi_{j} = \varepsilon_{jlk}\partial_{l}\Psi_{k}\ ,
\end{equation}
where $\Psi_{k}$ are the tree components of the photon state in the
tree dimensional Hilbert space for spin 1. (However, the photon
states are restricted to a two dimensional subspace spanned by the
helicity $\pm1$ states.)

Notice that one can \emph{formally} derive Maxwell's equations in
vacuum (that is, with vanishing charge and currents) from
Eq.(\ref{QMequPhoton}); however their interpretation is quite
different: the equation above describes the dynamics of a one photon
system whereas Maxwell's  equations are the evolution equations for
a set of observables --the electromagnetic field-- of a system made
by an indefinite number of photons. Furthermore the alleged
derivation is based on the conceptual error of identifying the tree
dimensional Hilbert space of spin with the tree dimensional physical
space\cite{dltPhoton}.

\section{Quantum dynamics in two dimensional space-time}
For Schr\"{o}dinger and Klein-Gordon equations, the specialization for
two dimensional space-time $(t,x)$ is straightforward: we just make
the association $E\rightarrow i\partial_{t}$ and $P\rightarrow
-i\partial_{x}$ and we get
\begin{equation}\label{SchroEq}
  \partial_{t} \psi(x,t) = \frac{i}{2m} \partial^{2}_{x}
\psi(x,t) \ ,
\end{equation}
\begin{equation}\label{KGEq}
     (\partial^{2}_{t}-\partial^{2}_{x})\psi(x,t)=-m^{2}\psi(x,t)\ .
\end{equation}
For Dirac equation we choose the two dimensional matrices and spinor
\begin{equation}\label{gama2D}
   \gamma^{0}=\left(%
\begin{array}{cc}
  0 & 1 \\
  1 & 0 \\
\end{array}%
\right)
  \ ,
   \gamma^{1}=\left(%
\begin{array}{cc}
  0 & 1 \\
  -1 & 0 \\
\end{array}%
\right)\ , \psi_= \left(%
\begin{array}{c}
 \psi_{1} \\
 \psi_{2} \\
\end{array}%
\right)
\end{equation}
and we become from Eq.(\ref{epDirac})
\begin{eqnarray} \label{dirac2D}
\partial_{t}\psi_{2}(x,t) &=&-\partial_{x}\psi_{2}(x,t)-i\,m\,\psi_{1}(x,t)
\nonumber \\
\partial_{t}\psi_{1}(x,t) &=& \partial_{x}\psi_{1}(x,t)-i\,m\,\psi_{2}(x,t)
\ .
\end{eqnarray}

The quantum equation for the photon in two dimensional space-time is
not trivial because the transversality constraint (\ref{transv}) and
equations (\ref{epMw}) and (\ref{QMequPhoton}) are essentially in
tree dimensional space. We can however make some formal
modifications of Eq.(\ref{QMequPhoton}) in order to get a two
dimensional equation. Let us fix the index $l$ at the value $l=1$
and associate $\partial_{1}=\partial_{x}$. The indices $j,k$ cant
take the values $2,3$ and $3,2$ resulting in two coupled equations:
\begin{eqnarray}\label{2dimQMequPhoton}
  i\partial_{t}\Psi_{2} &=& -\partial_{x}\Psi_{3}\ , \nonumber \\
   i\partial_{t}\Psi_{3} &=& +\partial_{x}\Psi_{2}\ .
\end{eqnarray}

Although the identification of Eq.(\ref{QMequPhoton}) with Maxwell's
equations is questionable, but formally possible, in the two
dimensional case we can also obtain the two dimensional Maxwell's
equations in vacuum with the misidentification of the photon states
with the electric and magnetic fields: $\Psi_{2}\rightarrow E$ and
$i\Psi_{3}\rightarrow B$  obtaining
\begin{eqnarray} \label{2dimMxwequ}
 \partial_{t}E &=& \partial_{x}B\ ,  \nonumber\\
 \partial_{t}B &=& \partial_{x}E\ .
\end{eqnarray}
These equations are also obtained by generalizing the four
dimensional Maxwell's equations to arbitrary dimension with the
techniques of differential geometry and then specializing it to two
dimensional space-time\cite{2dimelectro}.

The set of equations (\ref{2dimQMequPhoton}) can be decoupled with
the replacement $\Psi_{R}=\Psi_{2}-i\Psi_{3}$ and
$\Psi_{L}=\Psi_{2}+i\Psi_{3}$ obtaining
\begin{eqnarray}\label{2dimQMequRLphot}
   \partial_{t}\Psi_{R} &=& -\partial_{x}\Psi_{R}\ , \nonumber \\
  \partial_{t}\Psi_{L} &=& +\partial_{x}\Psi_{L}\ .
\end{eqnarray}
If we try solutions with the form $f(x-vt)$ we can see that
$\Psi_{R}$ and $\Psi_{L}$ represent a photon moving to the right
($v=1$) or to the left ($v=-1$) and therefore they are the one
dimensional counterparts of the tree dimensional right or left
handed photons. Notice that the $+-$ signs in
Eq.(\ref{2dimQMequPhoton}) correspond to different helicity states
and in Eq.(\ref{2dimQMequRLphot}) they correspond to different
directions of propagation.
\section{Schr\"{o}dinger Process}
Let us assume two types of virtual particles $A$ and $B$ in a one
dimensional lattice with sites at distance $\lambda$ . These
particles have an asymmetric interaction (violating the
action-reaction principle):  $A$ particles \emph{reject} $B$ from
its location whereas $B$ particles \emph{attract} $A$ to its
location. Metaphorically we can say that \emph{A hates B} and
\emph{B loves A} or that $B$ particles are \emph{good} and $A$ are
\emph{evil}. The rejection of a particle $A$ or $B$ from a site is
formalized by the creation of an antiparticle $\bar{A}$ or $\bar{B}$
at the site. Each site can be occupied by any number of particles of
type $A$ $B$ or by their corresponding antiparticles $\bar{A}$
$\bar{B}$. Particles and antiparticles of the same type annihilate
in each site of the lattice leaving only the remaining excess of
particles or antiparticles of both types $A$ and $B$. At each
discretized time step, $t\rightarrow t+1$, corresponding to a time
evolution by a small amount $\tau$, each particle of type $A$
creates two antiparticles $\bar{B}$ in the same site and one
particle $B$ in each of the two neighbouring sites. This is
equivalent to moving one particle $B$ to the right and another to
the left. In a similar way, $B$ particles move neighbouring $A$
particles attracting them to its site. This process is realized with
some amplitude (probability, if normalized) $k$.
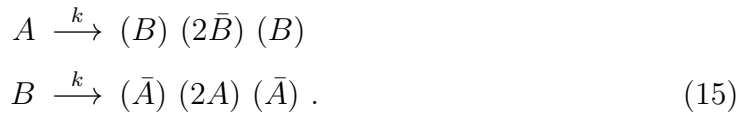
\begin{eqnarray} \label{SchrProc}
A &\stackrel{k}{\longrightarrow}&  (B)\  (2\bar{B})\  (B)\ \
\nonumber \\
B &\stackrel{k}{\longrightarrow}&   (\bar{A})\ (2A)\  (\bar{A})\ .
\end{eqnarray}
The same reactions occur exchanging particles and antiparticles.

Before we write the master equation for the time evolution, we can
notice some qualitative features of the process. If $A$ rejects $B$
but $B$ attracts $A$ then, transitively, $A$ reject themselves and
diffuse. Same conclusion is reached for $B$. It is obvious that the
process has diffusion because in each time step particles and
antiparticles are created in neighbouring sites. For instance, if we
start with two particles, one $A$ and one $B$, in one site, after
three time steps we will have 110 particles occupying seven sites.
It is less obvious that, even though the process has left-right
symmetry, we may also have drift to the right or to the left. In
order to see how this is possible, recall that $A$ particles {\em
reject} $B$ particles and $B$ {\em attract} neighbouring $A$
particles. Therefore, if we have an \emph{asymmetric} configuration
like $AB$, the center of the combined distribution will move towards
$B$. The drift direction and velocity is then {\em encoded} in the
shape and relative distribution of both types of particles. It is
remarkable that, although the distribution of particles are widely
distorted after few time steps, the drift direction and velocity
remain invariant (in the continuous limit).

Let $a_{s}(t)$ and $b_{s}(t)$ be the number of particles of type $A$
and $B$ respectively at the site $s$ at time $t$. When $a_{s}(t)$ or
$b_{s}(t)$ take negative values they denote the number of {\em
antiparticles}. At a particular site of the lattice, the number of
particles change as particles or antiparticles are created there by
the particles in neighbouring sites. The time evolution of the
process is then defined by the master equations
\begin{eqnarray} \label{MastEqScrProc}
a_{s}(t+1) &=& a_{s}(t)-k\;(b_{s-1}(t) -2b_{s}(t) + b_{s+1}(t))
\nonumber \\
b_{s}(t+1) &=& b_{s}(t)+ k\;(a_{s-1}(t) -2a_{s}(t) + a_{s+1}(t)) \ .
\end{eqnarray}
Choosing time and space scales such that $\frac{\tau}{\lambda^{2}}
=1$ we can write these equations as
\begin{eqnarray} \label{MastEqScrProc1}
 \frac{ a_{s}(t+1)-a_{s}(t)}{\tau} &=& -k\;\frac{ b_{s-1}(t) -2b_{s}(t) + b_{s+1}(t) }{\lambda^{2}}
\nonumber \\
\frac{b_{s}(t+1)- b_{s}(t)}{\tau} &=&  k\;\frac{ a_{s-1}(t)
-2a_{s}(t) + a_{s+1}(t) }{\lambda^{2}} \ .
\end{eqnarray}
In these equations one can easily recognize the discrete version of
the time derivative and of the second space derivative. Taking then
the limit $\tau\rightarrow 0$ and $\lambda\rightarrow 0$ with
$\frac{\tau}{\lambda^{2}}\rightarrow 1$ and $k=\frac{1}{2m}$ and
also replacing $a_{s}(t)$ and $b_{s}(t)$ by continuous functions
$a(x,t)$ and $b(x,t)$, the equations above become
\begin{eqnarray}\label{MastEqScrProc2}
\partial_{t} a(x,t) &=& -\frac{1}{2m}\;\partial^{2}_{x}
b(x,t)
\nonumber \\
\partial_{t} b(x,t) &=&  \frac{1}{2m}\;\partial^{2}_{x}
a(x,t) \ .
\end{eqnarray}
Now we can combine the real amplitudes of the virtual particles in
one complex field $\psi(x,t)= a(x,t) +ib(x,t)$ and both equations
above result in Schr\"{o}dinger equation given in Eq.(\ref{SchroEq}).

The virtual particles process presented here is a much simplified
version of a similar process studied in more detaills\cite{dltDal}.
\section{Dirac Process}
Let us assume now four types of virtual particles $A,B,C,D$ in a one
dimensional space. These particles always move and create and
annihilate each other. Precisely, $A$ and $B$ always move to the
left whereas $C$ and $D$ move to the right; $D$ particles create $A$
but $A$ annihilates $D$ and similarly,  $B$ particles create $C$ but
$C$ annihilates $B$.

Notice that with regard of the movement, the pair $(AB)$ behaves as
chirality opposed to the pair $(CD)$, that is, we may think of
$(AB)$ as having negative helicity (left handed) and $(CD)$ with
positive helicity (right handed). Furthermore, with respect with
creation and annihilation, the pair $(AC)$ behaves as the
antiparticles of the pair $(BD)$. Therefore the process globally
contains the essence of antimatter but not in an identified way,
that is, none of the particles is individually the antiparticle of
another. Another interesting qualitative feature is that since $D$
creates $A$ but $A$ annihilates $D$, transitively, $D$ has self
annihilation and the same can be said for all other particles. This
self annihilation component is essential to the virtuality of the
particle: they can not have permanent existence.

In order to formalize this process let us assume again an infinite
one dimensional lattice with sites at distance $\lambda$ occupied by
some number of these particles.  Each site can be occupied by any
number of particles of type $A,B,C,D$. At each discretized time
step, $t\rightarrow t+1$, corresponding to a time evolution by a
small amount $\tau$, particles move one lattice site and are created
and destroyed with some amplitude $k$. In this case we choose
space-time units and scale such that $\frac{\tau}{\lambda}=1$.

Let $a_{s}(t),\ b_{s}(t),\ c_{s}(t),\ d_{s}(t)$ be the number of
particles of type $A,\ B,\ C,\ D$ respectively at the site $s$ at
time $t$.  At a particular site of the lattice, the number of
particles change as particles migrate, are created or destroyed. The
time evolution of the process is then defined by the equations
\begin{eqnarray}\label{MastEqDirProc}
a_{s}(t+1) &=& a_{s+1}(t)+ k d_{s}(t) \nonumber \\
b_{s}(t+1) &=& b_{s+1}(t)- k c_{s}(t) \nonumber \\
c_{s}(t+1) &=& c_{s-1}(t)+ k b_{s}(t) \nonumber \\
d_{s}(t+1) &=& d_{s-1}(t)- k a_{s}(t)  \ .
\end{eqnarray}
We will see that this simple process is described by Dirac equation.
Let us first subtract the same quantity from each side of these
equations and then use the relation $\frac{\tau}{\lambda}=1$ to
obtain
\begin{eqnarray}\label{MastEqDirProc1}
\frac{a_{s}(t+1) - a_{s}(t)}{\tau}&=& \frac{a_{s+1}(t)- a_{s}(t)}{\lambda}+ \frac{k}{\lambda} d_{s}(t) \nonumber \\
\frac{b_{s}(t+1) -b_{s}(t)}{\tau}&=& \frac{b_{s+1}(t)-b_{s}(t)}{\lambda}-  \frac{k}{\lambda} c_{s}(t) \nonumber \\
\frac{c_{s}(t+1) -c_{s}(t)}{\tau}&=& - \frac{c_{s}(t)-c_{s-1}(t)}{\lambda} +  \frac{k}{\lambda} b_{s}(t) \nonumber \\
\frac{d_{s}(t+1) -d_{s}(t)}{\tau}&=& -
\frac{d_{s}(t)-d_{s-1}(t)}{\lambda} -  \frac{k}{\lambda} a_{s}(t) \
.
\end{eqnarray}

In these master equations one can easily recognize the discrete
version of the space and time derivatives. Taking then the limit
$\tau\rightarrow 0\ ,\ \lambda\rightarrow 0 \ ,\  k\rightarrow 0$
with $\frac{\tau}{\lambda}\rightarrow 1\ ,\
\frac{k}{\lambda}\rightarrow m$ and replacing
$a_{s}(t),b_{s}(t),c_{s}(t),d_{s}(t)$ by continuous functions
$a(x,t),b(x,t),c(x,t),d(x,t)$, the equations above become
\begin{eqnarray}
\partial_{t} a(x,t) &=& \partial_{x}a(x,t)+m\,d(x,t)
\nonumber \\
\partial_{t} b(x,t) &=& \partial_{x}b(x,t)-m\,c(x,t)
\nonumber \\
\partial_{t} c(x,t) &=&- \partial_{x}c(x,t)+m\, b(x,t)
\nonumber \\
\partial_{t} d(x,t) &=& -\partial_{x}d(x,t)-m\, a(x,t)
\ .
\end{eqnarray}
We can now  combine the first two real equations in a complex one by
adding the first with the second multiplied by $i$ (and similarly
for the last two)
\begin{eqnarray}
\partial_{t}\left( a(x,t)+i\,b(x,t)\right) &=& \partial_{x}\left( a(x,t)+i\,b(x,t)\right)-i\,m\,\left( c(x,t)+i\,d(x,t)\right)
\nonumber \\
\partial_{t}\left( c(x,t)+i\,d(x,t)\right) &=& -\partial_{x}\left( c(x,t)+i\,d(x,t)\right))-i\,m\,\left(  a(x,t)+i\,b(x,t)\right)
\ ,
\end{eqnarray}
and in terms of the complex fields $\psi_{1}(x,t)=a(x,t)+i\,b(x,t)$
and $\psi_{2}(x,t)=c(x,t)+i\,d(x,t)$, the expression above results
in Dirac equation given in (\ref{dirac2D}).

Notice that $\psi_{1}$ and $\psi_{2}$ have opposed helicity and
represent the antimatter of each other.

\section{Klein-Gordon Process}
After giving two stochastic processes with mean field equations
corresponding with Schr\"{o}dinger and Dirac equations we present now a
virtual particle process that results in Klein-Gordon equation
(\ref{KGEq}) arising from the relativistic energy-momentum relation
(\ref{epKG}).

The second space derivative is obtained in a similar way as in the
Schr\"{o}dinger process in Eq.(\ref{SchrProc}) (but with just one type
of diffusing particle) and the rest energy term is also simple, but
the \emph{second} time derivative presents some difficulty because,
in the discrete time version, it involves \emph{tree} times:
$t+1,\quad t,\quad t-1$. Notice that the Schr\"{o}dinger and Dirac
processes are markovian because the state at time $t+1$, given by
the master equations Eqs.(\ref{MastEqScrProc},\ref{MastEqDirProc}),
are uniquely determined by the state at time $t$, without memory on
past times. In the present case we must abandon the markovian
property because the master equation will also involve the state at
time $t-1$.

Let $a_{s}(t)$ denote the number of particles (antiparticles, if
negative) of type $A$ at site $s$ at time $t$ in a one dimensional
lattice with lattice constant $\lambda$. At each step $\tau$ of
discretized time $t\rightarrow t+1$ all particles reproduce and
migrate to neighbouring sites: $\circ A\circ \rightarrow A\circ A$
and are annihilated with probability $k$. The non-markovian property
is that all particles present at time $t-1$ are removed. The master
equation is then
\begin{equation}\label{MasEqKG}
 a_{s}(t+1) = a_{s+1}(t)+a_{s-1}(t) -k a_{s}(t) - a_{s}(t-1)\ .
\end{equation}
With some trivial manipulations, and using
$\frac{\tau^{2}}{\lambda^{2}}=1$, we can write this equation as
\begin{equation}\label{MasEqKGmod}
 \frac{ a_{s}(t-1)-2a_{s}(t)+a_{s}(t+1)}{\tau^{2}} =
 \frac{ a_{s-1}(t)-2a_{s}(t)+a_{s+1}(t)}{\lambda^{2}} -\frac{k}{\lambda^{2}} a_{s}(t)
\end{equation}
 Now, taking the limit $\tau\rightarrow 0\ ,\ \lambda\rightarrow 0 \ ,\  k\rightarrow 0$
with $\frac{\tau}{\lambda}\rightarrow 1\ ,\
\frac{k}{\lambda^{2}}\rightarrow m^{2}$ and replacing $a_{s}(t)$ by
the continuous field $\psi(x,t)$ we obtain Klein-Gordon equation
(\ref{KGEq}).

It is possible to recover the markovian property for a Klein-Gordon
process if we make an additional postulate. A process described by
$a_{s}(t)$ can be said to be \emph{causal or markovian} if the state
at $t+1$ is uniquely determined by the state at time $t$, that is,
it satisfies an evolution equation of the type
$a_{s}(t+1)=F[a_{r}(t)]$. Now we can define the process to have
\emph{reversed causality or reverse markovian} if the state at time
$t-1$, that is, not at future but at past time, is uniquely
determined by the state at time $t$. Such a process satisfies an
equation of the type $a_{s}(t-1)=F[a_{r}(t)]$. Notice that the
requirement of reversed causality is \emph{different }from time
reversal symmetry $T: t\rightarrow -t$. Causality means that the
\emph{future} is uniquely determined by the present whereas in
reverse causality the \emph{past} is uniquely determined by the
present. Let us consider then a markovian Klein-Gordon process
described by
\begin{equation}\label{MasEqKG1}
 a_{s}(t+1) =\frac{1}{2}a_{s+1}(t)+\frac{1}{2}a_{s-1}(t) -k\frac{1}{2}a_{s}(t) \ .
\end{equation}
This is a process similar to the one of Eq.(\ref{MasEqKG}) but
\emph{without} the non-markovian term $a_{s}(t-1)$. Now, if we
require reversed causality we also have
\begin{equation}\label{MasEqKG1}
 a_{s}(t-1) =\frac{1}{2}a_{s+1}(t)+\frac{1}{2}a_{s-1}(t) -k\frac{1}{2}a_{s}(t) \ .
\end{equation}
Adding the last two equations we get
\begin{equation}\label{MasEqKG3}
 a_{s}(t+1)+ a_{s}(t-1)= a_{s+1}(t)+ a_{s-1}(t) -k a_{s}(t) \ .
\end{equation}
This is the same master equation (\ref{MasEqKG}) for the
non-markovian process that in the continuous limit results in
Klein-Gordon equation.

Here we have taken a real field but the same could be done with a
complex or a many component field in order to account for particles
with charges.
\section{Photon Process}
The photon process is the simplest of all. Let us assume two types
of particles $R$ and $L$. In each time step $t\rightarrow t+1$ of
discretized time, corresponding to a time evolution by a small
amount $\tau$, $R$ particles move one site to the right and $L$
particles move to the left in a one dimensional lattice with lattice
constant $\lambda$. The particles can not be at rest and move with
constant speed $\frac{\lambda}{\tau}$. Let $r_{s}(t)$ and $l_{s}(t)$
denote the number of right and left moving particles at site $s$ at
time $t$. Then, the master equations are simply
\begin{eqnarray}
r_{s}(t+1) &=&r_{s-1}(t)
\nonumber \\
l_{s}(t+1) &=& l_{s+1}(t) \ .
\end{eqnarray}
Subtracting $r_{s}(t)$ [$l_{s}(t)$] on both sides of the first
[second] equation and choosing time and space scales such that
$\frac{\lambda}{\tau} =1$, we can write these equations as
\begin{eqnarray}
 \frac{ r_{s}(t+1)-r_{s}(t)}{\tau} &=& -\frac{ r_{s}(t) -r_{s-1}(t)}{\lambda}
\nonumber \\
\frac{l_{s}(t+1)- l_{s}(t)}{\tau} &=&  \frac{l_{s+1}(t) -l_{s}(t)
}{\lambda} \ .
\end{eqnarray}
Here we find the discrete version of time and space derivatives.
Taking then the limit $\tau\rightarrow 0$ and $\lambda\rightarrow 0$
with $\frac{\tau}{\lambda}\rightarrow 1$ and replacing $r_{s}(t)$
and $l_{s}(t)$ by continuous functions $\Psi_{R}(x,t)$ and
$\Psi_{L}(x,t)$ we obtain the quantum mechanic dynamic equation for
a photon given in Eq.(\ref{2dimQMequRLphot}).

\section{Conclusion}
In a way consistent with the quantum field theory interpretation of
quantum mechanics we have found stochastic processes with virtual
particles whose mean field equations correspond to Schr\"{o}dinger,
Dirac, and Klein-Gordon equations, as well as the corresponding
quantum mechanic equation for the photon.

One possible interest of studying these ``quantum stochastic
processes'' appears if we consider that the mean field equations are
obtained when we take the limit from a discrete to a continuous
space-time. However it is well known in non-equilibrium statistical
mechanics that there are some features of discrete space-time
processes that could appear in a computer simulation that are not
present in the solutions of the mean field equations. This could be
interesting in order to predict some effects due to an inherent
discreetness of space-time of the order of Planck time $T_{P}\approx
10^{-42}s$ and length $L_{P}\approx 10^{-35}m$ as is suggested in
some quantum gravity theories. Such simulations with $\tau \approx
L_{P}$ and $\lambda\approx L_{T}$ are beyond reach with today
computing power, however with the exponentially increasing computing
capabilities it may sometime be possible.

I would like to thank ANSES for financial support and M. Hoyuelos
and H. M\'{a}rtin for their comments.

\end{document}